\documentclass{article}

\usepackage{arxiv}

\usepackage[utf8]{inputenc} 
\usepackage[T1]{fontenc}    
\usepackage{hyperref}       
\usepackage{url}            
\usepackage{booktabs}       
\usepackage{amsfonts}       
\usepackage{nicefrac}       
\usepackage{microtype}      
\usepackage{lipsum}
\usepackage{graphicx}
\graphicspath{ {./images/} }

\title{Experimentation in Gaming: an Adoption Guide 
}

\author{
 Julian Runge \\
  Northwestern University\\
  Medill School of Journalism, Media, Integrated Marketing Communications\\
  1845 Sheridan Road; Evanston, IL 60208 \\
  \texttt{julian.runge@northwestern.edu} \\
  \\
   \And
 Draft for feedback.\\
 This article is to be published on the \textit{Outperform} learning platform and in the H1/2025 magazine.\\
 }

\begin{document}
\maketitle
\begin{abstract}
Experimentation is a cornerstone of successful game development and live operations, enabling teams to optimize player engagement, retention, and monetization. This article provides a comprehensive guide to implementing experimentation in gaming, structured around the game development lifecycle and the marketing mix. From pre-launch concept testing and prototyping to post-launch personalization and LiveOps, experimentation plays a pivotal role in driving innovation and adapting game experiences to diverse player preferences. Gaming presents unique challenges, such as highly engaged communities, complex interactive systems, and highly heterogeneous and evolving player behaviors, which require tailored approaches to experimentation. The article emphasizes the importance of collaborative frameworks across product, marketing, and analytics teams and provides practical guidance to game makers how to adopt experimentation successfully. It also addresses ethical considerations like fairness and player autonomy.
\end{abstract}


\section{Introduction}
At the core, in gaming as elsewhere, experimentation and AB testing are about learning from customers, both free and paid, to identify more optimal designs and strategies that drive relevant performance indicators such as engagement, retention, monetization and user satisfaction. A successful experimentation program needs a roadmap for hypothesis-driven learning and innovation, centrally ensured rigor in experiment design and analysis, mechanisms for knowledge accumulation and dissemination \cite{HBR2025}, executive endorsement and a culture that allows for risk-taking and failure \cite{HBR2020}.

Gaming, however, offers up unique challenges and opportunities in experimentation. Important differences exist along different stages of the game development and publishing process. Both pre- and post-launch, experimentation informs important steps for bringing a game to market and managing it. While emergent game creation is the focus during earlier stages, commercial optimization becomes more important as a game is live in the market. With game designers, there can be concerns that experiment-driven optimization may compromise the artform that is game design – something that needs to be carefully managed through cross-departmental collaboration and a shared learning agenda.

There are further aspects that are unique to experimentation in gaming: Highly engaged player communities can make it hard to run certain types of experiments, e.g., with different prices \cite{RungeDeconstructorFun}, matchmaking algorithms \cite{RungeInformationSystems}, or other variations that might be perceived as unfair. As games are highly immersive and interactive experiences, the set of relevant indicators can be larger and more diverse \cite{SinghSegWise2024} than in other areas of experimentation. The publishing model (premium versus free-to-play), platform strategy (mobile, console, PC, cross-platform), growth trajectory (paid and/or organic), and ongoing development tactics (LiveOps, content updates) impact to what extent and how you can leverage experimentation. In selecting your game’s technology stack, you should also consider what experimentation and personalization strategies you want to support \cite{RungeMedium2019} once the game is live in market.

This article presents an introduction to touchpoints between game making and experimentation, structured against a matrix of a four-stage game development and publishing process and the four Ps of the marketing mix (product, place, price, promotion). To get started, let’s develop this conceptual matrix and populate it with key tasks essential to getting a game to market. We will then discuss how these tasks interact with experimentation, highlight key challenges and opportunities, and outline pathways that will help you succeed with experimentation in gaming.

\section{The Game Development and Publishing Process}

A 2023 blog post by Egor Piskunov \cite{PiskunovILogos2023}, an experienced game developer, offers up a detailed account with eight essential stages. For our purposes, let’s simplify it a bit to four stages:

\begin{figure} [h]
    \centering
    \includegraphics[width=0.8\linewidth]{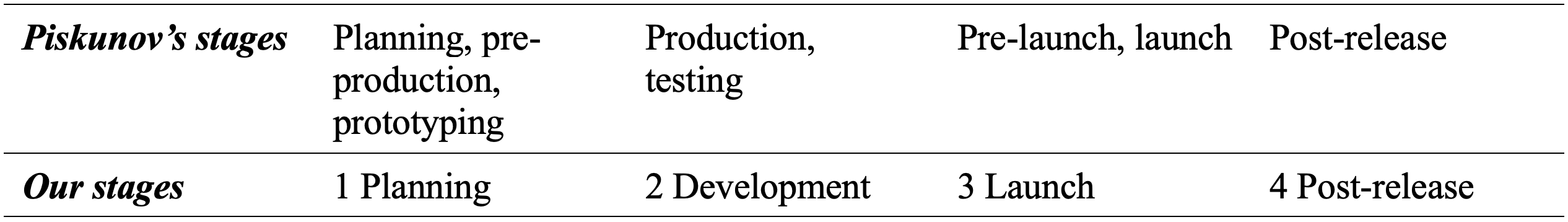}
  
\end{figure}

As noted, experimentation in digital media and interactive entertainment is about learning from consumers, i.e. players. Our conceptualization hence consolidates earlier consumer-distant stages while later in-market stages receive comparatively more attention.

To identify the areas where experimentation can drive value at each stage, let’s lean into a concept that captures the strategy set available to companies when interacting with consumers: the marketing mix with its 4P \cite{BordenMarketingMix} product, place (aka distribution), price, and promotion. By combining our game lifecycle with the marketing mix, we obtain a matrix that we can populate with key tasks involved in bringing a game idea to market:

\begin{figure} [h]
    \centering
    \includegraphics[width=0.8\linewidth]{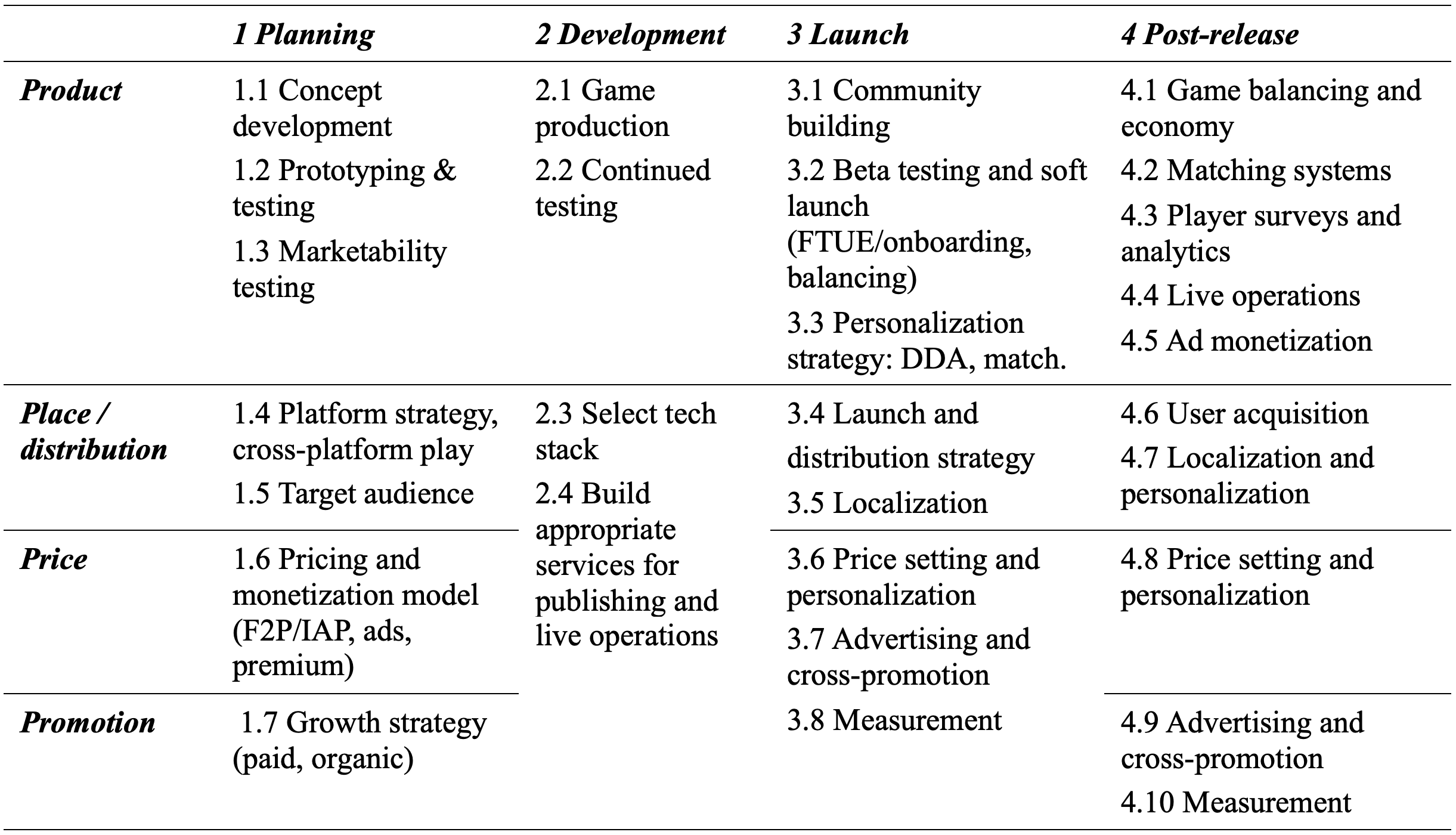}

\end{figure}

To get on the same page about what experimentation is: It exposes different versions of a user experience – from button colors over difficulty adaptation systems \cite{CBSRunge} and promotional strategies \cite{RungeSpringerNatureLink2022} to whole product features – to customers to learn about their reaction. Pre-launch, experimentation tends to be \textit{qualitative} and occur on small samples, often without randomization, and with mostly qualitative performance measurement, e.g., through player or expert feedback (is it fun?). During soft launch (i.e., the launch of the game in select test markets), sample sizes grow, randomization becomes a crucial ingredient, and qualitative and quantitative performance measurement tend to both be important. When the game is launched, with live games and large audiences, the emphasis shifts noticeably towards \textit{quantitative experimentation} with large samples, randomization, and quantitative performance measurement.

So now, let’s use our matrix to highlight key tasks that interact significantly with qualitative (so, the pre-launch type with small samples) and quantitative (so, the launch and post-release type with large samples) experimentation, or both. The following matrix will be the backbone for our further discussion:

\begin{figure} [h]
    \centering
    \includegraphics[width=0.8\linewidth]{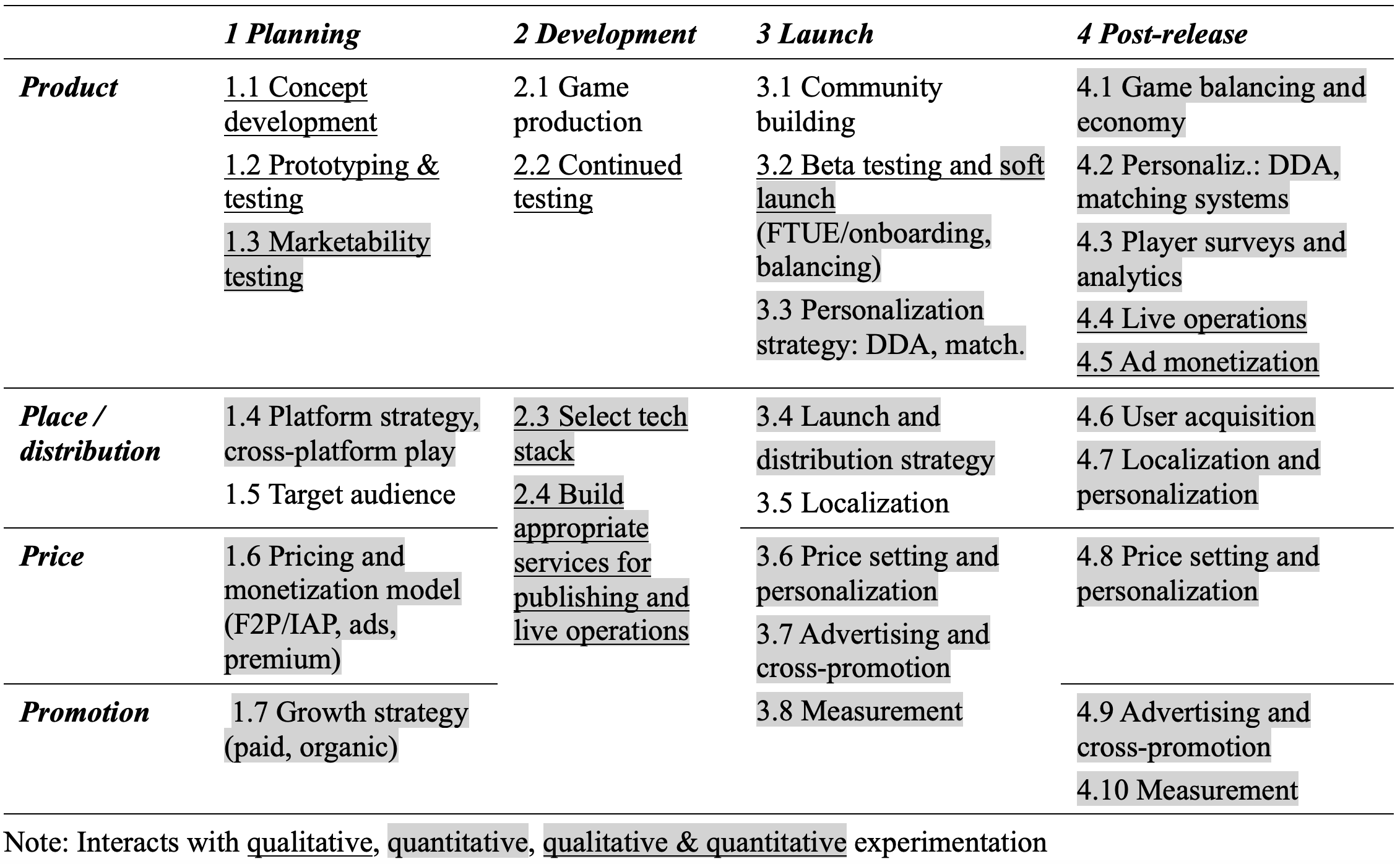}
\end{figure}

\section{Experimentation Before Launch (Stages 1-3)}

Working from left to right in our matrix, the pricing and monetization model you choose for your game (1.6) and the intended growth strategy (1.7) will have major impact on your ability to leverage experimentation. Freemium (F2P, free-to-play) games with strong organic growth tend to attract larger audiences, in turn providing a stronger basis and larger sample sizes for experimentation. Monetization design elements such as lootboxes and high-price IAPs (in-app purchases) that skew spending distributions will impact the variance you face in measuring experimental effects. Similarly, if you plan for a paid growth strategy with high-value players and heavy monetization mechanics, you will likely have a (much) smaller player base, impacting your ability to run quantitative experiments. If you also plan to monetize via ads, you may need to run more experiments to optimize overall monetization. If you publish a premium game that players need to buy upfront, you will mostly care about player engagement and retention metrics (and maybe ad monetization).

A type of experimentation, sometimes called pre-release experimentation \cite{LinakerBjarnasonFagerholmAriX}, often happens in user testing during the game planning and development phases (1.1, 1.2, and 2.2). This type of experimentation is mostly qualitative and hence quite different from experimentation and AB testing on end users and at scale. Sample sizes are much smaller. Examples for pre-release experiments \cite{BjarnasonSpringerNatureLink2024} involve having test users and experts play different prototypes or inviting mock reviews by game journalists and experts. To the extent that generative AI models to emulate in-market player preferences \cite{llms-market-research-2024} are available, experiments with prototypes (or later, release candidates in LiveOps, 4.4.) can also be run on AI agents. At a minimum, such experimentation can serve to identify bugs and technical problems \cite{MobileDevMemPodcast}.

A typical task in the planning stage that can involve large-scale experiments is marketability testing (1.3), e.g., of game theme and creative strategy. This involves releasing images and videos on social media or in digital advertising campaigns and experiments to gather customer feedback. The advanced ad testing capabilities of today’s large ad platforms \cite{RungeBusinessHorizons} can facilitate quantitative experimentation on large samples for this task.

When you decide on the technologies you will use to publish and distribute your game/s (2.3 and 2.4), you should keep an eye out for solutions that integrate well with AB testing and experimentation at scale. Developing backend services such that they can flexibly place players into different configurations and experiences \cite{RungeMedium2019} from the get-go can save substantial time and cost later on.

As your game develops and approaches launch, you will usually beta test and then soft launch the game in a few test markets before entering worldwide release. Many companies experiment heavily during these steps, both qualitatively and quantitatively, to verify important hypotheses on onboarding and FTUE (first-time user experience) and game difficulty and balancing (3.2). It is also the time to either experiment with or define the strategy for personalization in product (3.3) and pricing (3.6). Having clarity on the approach here can help you ready the systems to test DDA (dynamic difficulty adaptation) \cite{CBSRunge}, matching (3.3, 4.2) \cite{RungeInformationSystems} and price and offer targeting strategies (3.6, 4.8) \cite{RungeDeconstructorFun} quickly when you reach scale.

Soft launch can be the crucial stage to assess PMF (product-market fit) and make launch-or-kill decisions \cite{PaananenSuperCell}. As sample sizes are small and time is limited during soft launch, you should only experiment with changes that you expect to have major impact, e.g., drastically different FTUEs, balancing scenarios (3.2) and personalization strategies (3.3, 3.6), and keep the number of different experimental conditions to a minimum. This is not the time to look for precise quantification of treatment effects but to test major assumptions and design differences.

Pre-launch is also the time to get in gear to prepare experiments on key advertising and cross-promotion channels (3.7). Your launch and distribution strategy (3.4), e.g., location and budget, will impact what types of ad experiments you can expect to run. Be aware of the possibilities and limitations here and plan accordingly. These experiments can then serve to validate and calibrate your marketing measurement solutions (3.8) \cite{RungeGameDataPros}, e.g., media and marketing mix models \cite{HBR2023}. The experimentation strategy here will usually be developed by the company’s publishing organization, e.g., the marketing or user acquisition team. The product-centered experiment strategy (1.1-3.3), on the other hand, is commonly owned by the studio or game team. Our matrix above can serve to structure conversations and alignment what team should lead what experimentation use cases – more on this later.

\section{Experimentation After Launch (Stage 4)}

Post-release, as your game’s audience grows, experimentation really kicks into gear. You now have the player numbers (and hence sample sizes) to test hypotheses, design elements, and systems that remained untested pre-launch. Common applications are level and economy balancing \cite{CBSRunge}, reward systems \cite{WaikarGSB}, currency endowments and giveaways \cite{PLOSVirtualCurrency}, and algorithmic systems for adaptation and personalization of the user experience (4.1-4.2) \cite{RungeMobileDevMemo}. Precise and reliable tracking of player feedback and analytics (4.3) is a crucial ingredient to experimentation in gaming as it makes measurement of outcome metrics and diagnostic implementation checks \cite{DengsøeEppo} possible.

Live operations in digital gaming (4.4, LiveOps for short) refer to the continuous release of new content, e.g., new level packs, maps, or game events. LiveOps teams often operate on weekly schedules. Experimentation can be an important tool to inform well-balanced and -tailored LiveOps. However, the time pressure can be prohibitive for rigorous experimentation practices. Measurement windows tend to be short (with weekly events, maximum a week by definition) and a bias towards action can make it hard to spell out well-rounded hypotheses and set up well-designed experiments. In my experience, rigorous experiment design and analysis, e.g., enforced by a central platform \cite{LuchtEppo}, is crucially important for reliable decision support despite the pace of operations.

Additionally, I have sometimes observed in practice what can be dubbed the “measure-it-all” fallacy where LiveOps managers want to understand the impact of different actions in detail, ideally every week. This is a hard-to-achieve ambition where education by game analysts and data scientists is important for realistic expectations. Given the speed of iteration and focus on action, and depending on audience size, only a few quantitative experiments with short measurement windows will be possible for each content release. A structured experimentation roadmap and a knowledge repository summarizing design assumptions and hypotheses to be measured and results can help facilitate learning across event cycles and for the longer run. Test one thing at a time and improve the designs you release each week using a learning agenda and knowledge repository. Consider conducting a hackathon with experimentation data scientists and the LiveOps team \cite{SilvermanEppo} to set up the learning roadmap. And complement quantitative with qualitative experimentation to fill in gaps where you don’t have sample size or face other constraints.

Ad monetization (4.5) and price setting and personalization( 4.8) \cite{RungeCoG2024} are further areas where quantitative experimentation can drive much value by helping find revenue-optimal placements and incentive schedules (discounts or bonuses for the case of price, rewards for the case of incentivized in-game ads). Similarly, experimentation can inform what value localization and personalization initiatives provide (4.7), to decide which ones have merit and which ones do not.

Finally, core marketing efforts such as user acquisition (4.6), advertising and cross-promotion (4.9) \cite{RungeIEEE2014}, and measurement (4.10) can benefit significantly from quantitative experimentation, in particular through precise estimates of marketing effects. These precise estimates can precede major marketing investments and calibrate and validate observational analytics models such as media and marketing mix models (4.10) \cite{RungeGameDataPros}. 

For continuous LiveOps decision support (4.4), the analytics team could maintain a similarly calibrated observational model that continuously provides estimates of the effects of different actions. This observational model could be regularly and repeatedly calibrated using highly valid experimental results \cite{RungeGameDataPros} from the AB tests that the LiveOps team is able to run (see above, helping to “measure-it-all”). Analytics users, so LiveOps managers in this instance, should thereby understand that the observational estimates tend to have much larger risk of being wrong. This understanding can come from appropriate education of analytics users by analysts and data scientists.

\section{Who Should Own What Area of Experimentation?}

As mentioned in the beginning, success with experimentation in gaming requires carefully bridging the artform that is game design and analytical commercial aptitude. Operationally, this means that game design, product management, marketing, and analytics (incl. user research) need to work together smoothly and with clear direction. Game design (and art) develop content and systems, product management packages those for the market, marketing gets customers to engage with them, and analytics is essential to all three.

Product management and marketing are usually the business owners of experimentation, and analytics is the technological and methods owner. Our matrix can help you coordinate what areas of experimentation should be owned by product management, marketing, or \textit{both}; and what areas require deeper involvement by analytics and data science:

\begin{figure} [h]
    \centering
    \includegraphics[width=0.8\linewidth]{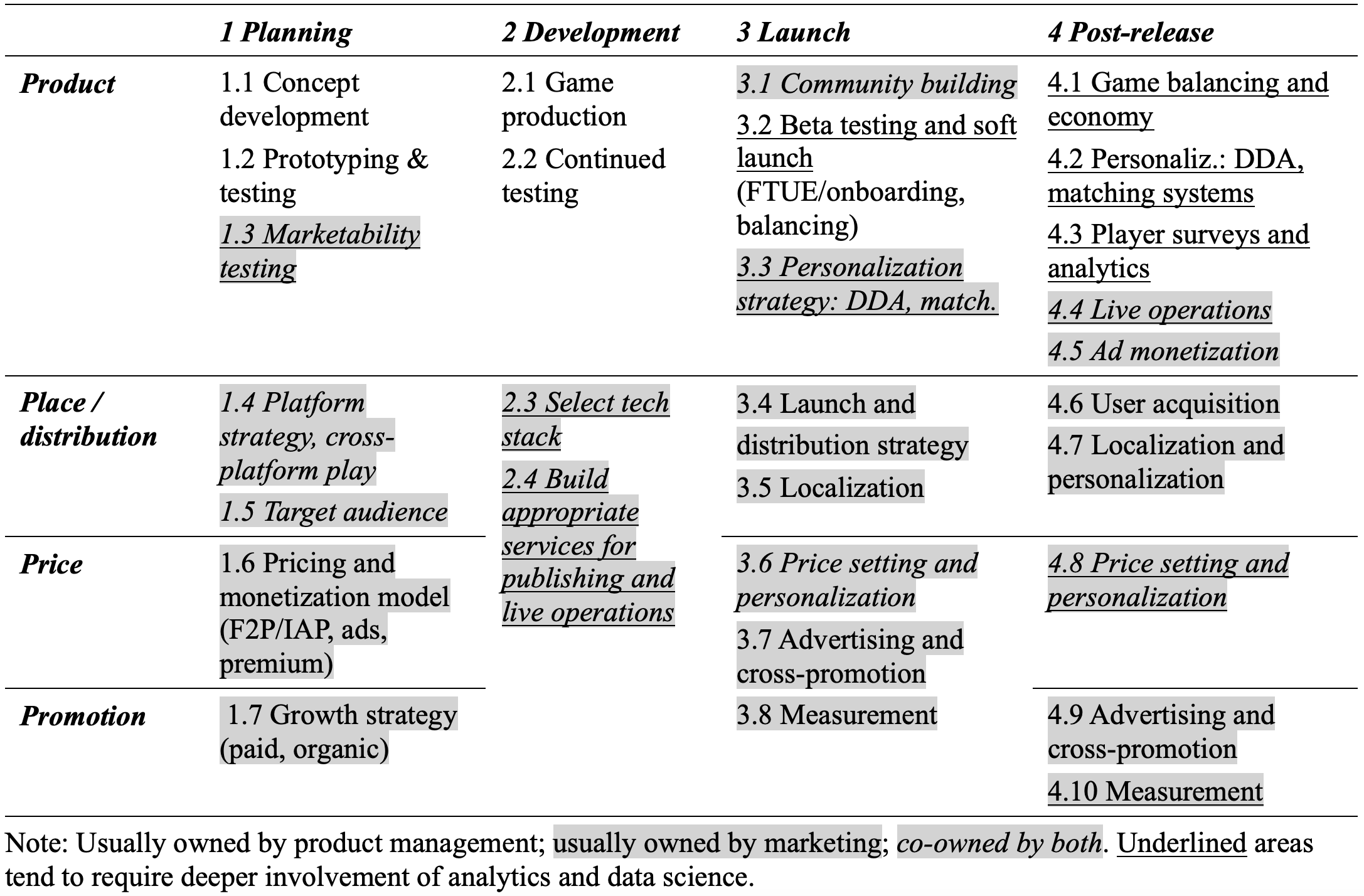}
\end{figure}

In this stylized representation of ownership structures, it is product management’s job to loop in game design as appropriate. In my experience, this looping-in of designers can be helpful across the full stages 1, 2, 3, and the product dimension of stage 4. It can even unlock high value for highly analytical exercises, e.g., to leverage the full richness of creative ideation when considering applications of algorithmic matching, targeting, and personalization systems (3.3, 3.6, 4.2, 4.7, 4.8).

The right ownership structure for experimentation also depends on the wider publishing model of the company. E.g., a large portfolio of hypercasual games will likely mean more control for marketing. A small portfolio with one or two games with deep and complex gameplay on the other hand will likely mean more control for product.

If LiveOps is a strong focus and revenue driver, product management and marketing may want to make specific arrangements for collaborating on weekly releases. This can take the form of standing sync meetings or personnel embedded in both teams to ensure messaging and theme of events and related marketing materials are aligned.

I have often seen success when certain areas involved all three experimentation owners, so product, marketing, and analytics. Those are marketability testing (1.3), tech and service delivery stack (2.3, 2.4), setting the personalization strategy (3.3), LiveOps (4.4), and price and offer personalization in the live game (4.8). Having all three stakeholders participate when you develop a strategy in these areas can help ensure you leverage the right external tools, methods, and treatments.

A key point of contention often arises in price personalization. While marketers want to leverage this important revenue optimization technique \cite{GrossoGameDataPros}, game teams may push back because of fairness and community-related concerns. A safe strategy I’ve often seen in practice is to personalize prices at the country-level within a reasonable corridor (e.g., a maximum factor of two between cheapest and most expensive pricing) and to use targeted offers for consumer-level personalization \cite{WaikarGSB} from there.

To develop effective offers for targeting in free-to-play games, product management should work with game designers to create bundles that truly drive value for players across different stages of the player journey in the game. Each bundle should come with a nice discount or bonus, ranging in magnitude from 30\% to a maximum 70\%. The bundles should cover different price points, after discount / bonus, from \$1.99 to \$99.99, and even \$199.99 for long-tail revenue games. I like to give higher discounts on smaller bundles as it can otherwise be hard to make small bundles have a clear value proposition (in terms of impact on the user experience). Experimentation can then test different targeting strategies \cite{RungeDeconstructorFun} such as skimming, country- and device-based, and RFM-style (recency, frequency, monetary value) personalization. A clear learning agenda synced across product, marketing, and analytics can help identify a close-to-optimal setup for the game step-by-step.

\section{Unique Characteristics of Live Games and How to Manage Them in Experimentation}

When you operate a live game, quantitative experimentation can truly make a difference. In my experience, it is important to be cognizant of four unique characteristics of gaming that present both challenges and opportunities in experimentation. Let’s dive in:

\subsection{Highly engaged communities}

\textit{Challenge}: Experiments in gaming often involve changes to core gameplay elements or the game’s balancing and economy, which can disrupt players’ experience. Players may notice differences between test groups, potentially leading to dissatisfaction or even churn if the changes are perceived as negative.

\textit{Opportunity}: Successful games have hugely engaged audiences with a strong sense of community and pervasive peer and social effects. Engaged audiences make it easier to detect subtle effects of experimental changes, enabling tests on nuanced design elements like pacing, narrative delivery, or even audio-visual cues. Matchmaking and messaging between players offer ample opportunity to test various mechanisms and approaches. 

\textit{What to do}:
\begin{itemize}
    \item Ensure that experiments do not break immersion or alienate players. Use subtle, incremental changes for testing.
    \item Carefully run tests that create noticeable discrepancies in user experiences on small test groups (e.g., in a marginal small market only) or in small-audience games to then transfer learnings to your games with large audiences.
\end{itemize}

\subsection{Complex immersive systems}

\textit{Challenge}: Games can be highly complex systems with many interdependent elements, such as mechanics, narratives, economies, and multiplayer dynamics. A change in one area often affects others, complicating the interpretation of test results. Understanding causal relationships in such intricate systems can be difficult as changes may have unintended ripple or delayed effects, making it challenging to capture their full impact within short time frames. For instance, adjustments to progression mechanics might influence long-term retention rather than immediate behavior.

\textit{Opportunity}: Games are immersive and self-contained. Players often interact with the game world in ways that can be tightly controlled by developers, reducing external confounding factors. This control can allow for precise isolation of variables in AB tests, such as how changing the spawn rate of rare items impacts player satisfaction. Many games feature in-game economies with virtual currencies, items, and other monetizable assets. These economies mimic real-world economic systems but operate in fully controlled environments, yielding immense amounts of rich data and making them ideal for testing pricing, scarcity effects, or promotional tactics.

\textit{What to do}:
\begin{itemize}
    \item To understand long-term effects, use predictive modeling to estimate long-term outcomes \cite{CutstomerLifetimeValuePrediction} based on short-term data, e.g., via surrogates like this study in a freemium app context \cite{TargetingforLong-term}, and complement AB tests with cohort analyses.
    \item Where useful, extend testing periods to understand long-term effects with certainty (e.g., see this experiment \cite{RungeSpringerNatureLink2022} that ran over nine months to evaluate different promotional strategies).
    \item Make use of advanced experiment methodologies like switchback experiments \cite{RungeInformationSystems} or cluster-level randomized designs (e.g., \href{https://facebookincubator.github.io/GeoLift/}{geo experiments}), especially when experimenting with peer effects and social systems that lead to network interference \cite{ZhuArXiv}.
\end{itemize}

\subsection{Diverse, heterogeneous audiences}

\textit{Challenge}: Gaming audiences tend to be highly diverse, with varying skill levels, preferences, and play styles. A change that improves the experience for one segment of players may detract from it for another. Experiment results may be skewed by player heterogeneity, leading to suboptimal decisions for specific segments. 

\textit{Opportunity}: Heterogeneity in play styles, skill, monetization behavior, and engagement patterns calls for personalization. Large amounts of granular and rich data are available from gameplay interactions, in-game purchasing, player progression, and social interactions. This data can be used for precise measurement of user behavior and experimental outcomes – and to personalize experiences. 

\textit{What to do}:
\begin{itemize}
    \item Always analyze treatment effects by important segments that capture heterogeneity in skill level, time in game, usage intensity, and spending patterns.
    \item Use personalization algorithms such as contextual bandits \cite{SchmitEppo} to tailor experiences to individual player profiles “on the fly.” Here’s a case study \cite{RungeCoG2024} for starter pack targeting, and here’s one for in-game creative \cite{RiekeGameDataPros}. 
    \item Have unified definitions for outcome metrics and heterogeneity covariates used in personalization, e.g., for retention, engagement, monetization, and social activities. These can power experimental analysis, predictive analytics, and (algorithmic) user experience personalization.
\end{itemize}

\subsection{Continuously evolving game content and player preferences}

\textit{Challenge: }New game content and features can lead to pervasive novelty effects, i.e., they excite players just because they are new. Early engagement indicators can then be severely misleading of long-term retention with the feature or design.

\textit{Opportunity}: Continuous development and the constant release of new game content and features, e.g., in LiveOps events, create ongoing opportunities for experimentation. Developers can deploy AB tests during these updates to refine mechanics, improve monetization, or enhance user experience.

\textit{What to do}:
\begin{itemize}
    \item Remain cognizant that novelty effects can be strong. Consider keeping a small longer-term holdout group to assess long-term effects with certainty and look at treatment effect heterogeneity in player tenure with the game (novelty effects tend to be lower or non-existent with new players).
    \item Have roadmaps and learning agendas across release cycles to develop a finely crafted, constantly improving game experience.
\end{itemize}

\section{Effective Innovation with Experimentation in Gaming}

To innovate effectively with experimentation in gaming, it is paramount that you learn over time as a game makes its way to and through the market and across games in your portfolio. Facilitate learning with a knowledge repository, collaborative learning agendas \cite{HBR2025}, and smart analytical frameworks \cite{RungeHumboldUniversity}. E.g., if you have a larger portfolio of games, experiment with more extreme treatments and disruptive changes in your smaller games and transfer learnings to the large games that generate the bulk of your revenue. This strategy mitigates the risk of upsetting or otherwise harming the communities of the largest games and allows for effective innovation across the portfolio of games.

When it comes to analytical frameworks, game economists have modeling frameworks \cite{VirtualEconomies2014} to design and balance economies for various types of games. An analytical tool I often make use of is what I call the Engagement Engineering framework \cite{RungeMobileDevMemo}. This framework draws on a general theory of human motivation, the basic needs concept within self-determination theory \cite{RyanDeci2008}. According to this concept, humans need to experience competence, relatedness, and autonomy to become intrinsically motivated for an activity. In leveraging this framework, I assume that strong long-term engagement is most sustainably achieved via \textit{intrinsic} player motivation. Accordingly, you should strive to maximize players’ experiences of competence, relatedness, and autonomy to obtain the highest levels of engagement and then work to optimize monetization of this engagement via ads and IAPs.

\subsection{Engineering Engagement from the Top Down}
For a top-down application of the framework \cite{RungeMobileDevMemo}, consider using it to develop innovations that drive competence, relatedness, or monetization and then check that your ideas do not interfere with players’ autonomy and free choice. Innovations that can fuel player feelings of competence are strong FTUEs, good tutorials, clearly set and communicated goals, e.g., via a continuous mission system, and systems for dynamic difficulty adaptation \cite{ZohaibDDA}. Especially with DDA systems though, apply the autonomy filter to check that the system will not unfairly impact players or foster addictive behaviors with certain player types (see the first case study here for autonomy-safe design of a DDA system for a puzzle game) \cite{RungeMobileDevMemo}. 

Innovative ideas that can serve to elevate feelings of relatedness are smart matchmaking, messaging, and other social systems like guilds. Successful approaches often group players together based on skill and engagement patterns (e.g., see the second case study here) \cite{RungeMobileDevMemo}. Also here, use the autonomy component of the Engagement Engineering framework to check that you don’t create player teams and groupings that are too similar within and too different between groups. Extreme approaches to matching and matchmaking can harm the long-term cohesion of your player community, serving to lower long-term engagement.

In monetization, offer personalization can drive a lot of value, and, if tuned appropriately, cater to the specific choice situations of players and delight them \cite{RungeAdvancedAnalytics}. If you however target low-price offers to players who would like to buy large bundles of in-game goods (or high-price offers to players who would like small bundles), you curtail their autonomy to purchase what they want, leading to unsatisfying experiences.

Similarly, fair and balanced lootboxes that reward players commensurate with overall pricing of in-game goods can make for an engaging and fun monetization system. Lootboxes, however, that trick players into sinking money into them in pursuit of one extremely rare (and largely elusive) high-powered item can harmfully interfere with feelings of autonomy and free choice. Especially players who are prone to overspend on lotteries and gambling may end up with unsatisfactory experiences, harming their long-term engagement. By using such monetization system, you are essentially at risk of trading a high-value gaming brand and long-term revenue and engagement for short-term revenue spikes.

\subsection{Engineering Engagement from the Bottom Up}
For a bottom-up application of the framework \cite{RungeMobileDevMemo}, pull analytics on how your players spend their time in the game, either overall or segmented by key player types (that can, e.g., be identified via clustering) \cite{DrachenArXiv}. Then assess if what you find aligns with what you expect and want. And think of ways how you can effectively increase engagement with the game given player behavior:

\begin{itemize}
    \item If players spend 70\% puzzling, is there anything you can do to increase feelings of competence for players as they puzzle? Hint: DDA \cite{CBSRunge}.
    \item If they spend 40\% messaging with other players in their teams or guilds, what can you do to increase experiences of relatedness during these interactions? Hint: matching systems \cite{RungeInformationSystems}.
    \item If players spend very little time in the game shop, what can you do to increase attention for the shop? Hint: create monetization need and desire \cite{ShengSellingBonus}, use popups and incentives to direct player attention \cite{RungeAdvancedAnalytics}.
\end{itemize}

There are many valid and useful analytical frameworks to support you in experimentation-based innovation in gaming. For accessible examples, consider Luton (2013) \cite{LutonFree-to-Play}, Lehdonvirta and Castronova (2014) \cite{VirtualEconomies2014}, and other works by these authors. For more advanced readings, consider the section on “Video game design” on \href{https://christopher-thomas-ryan.github.io/}{C. T. Ryan’s website}, e.g., \cite{LiOptimalSequencing} or \cite{LiOptimalDesign2023} with backward and forward citations.

The key is to have a behavioral and conceptual understanding of your players and the systems that bring them together \cite{RungeIndustrialOrganization} to anticipate where changes will have impact and why. On a technology level, it is important that you have an experimentation platform with a knowledge repository \cite{HBR2025} that allows for effective and efficient setup, analysis, and documentation of experiments.

\section{Time to Get Started}

The often long development processes for games and the unique nature of live games – with highly engaged communities, complex systems, diverse audiences, and continuous development – present both unique challenges and opportunities for experimentation. This article equips you with key knowledge to set up a successful experimentation practice. Doing so will enable you to create deeply engaging experiences, adapt to changing player preferences, and drive sustained growth in a highly competitive industry.

Current developments such as the proliferation of Generative AI \cite{RungeHarvardGamesRelationships} will impact the gaming industry, the game development process, and continuous release cycles in LiveOps significantly over time. Blending the abilities of Generative AI for content production with smart learning from customers through experimentation is bound to become a distinguishing competitive advantage.

Experimentation is here to stay and its future in gaming truly exciting.



\end{document}